\documentstyle[11pt,paspconf,epsf]{article}

\begin{document}

\def\avrg#1{\langle{#1}\rangle}
\newcommand{\ie}{{\em i.e.\ }}
\newcommand{\eg}{{\em e.g.,\ }}
\newcommand{\etal}{{\em et al.\ }}
\newcommand{\cf}{{\em cf.\ }}

\title{Astronomical Archives of the Future:\\ a Virtual Observatory}

\author{A.S. Szalay}
\affil{Department of Physics and Astronomy, The Johns Hopkins University,
Baltimore, MD, 21218}

\author{R.J. Brunner}
\affil{Department of Astronomy, California Institute of Technology, 
Pasadena, CA, 91125}

\begin{abstract}

Astronomy is entering a new era as multiple, large area, digital sky
surveys are in production. The resulting datasets are truly remarkable
in their own right; however, a revolutionary step arises in the
aggregation of complimentary multi-wavelength surveys (\ie the
cross-identification of a billion sources). Federating these different
datasets, however, is an extremely challenging task. With this task in
mind, we have identified several areas where community standardization can
provide enormous benefits in order to develop the techniques and
technologies necessary to solve the problems inherent in federating
these large databases, as well as the mining of the resultant
aggregate data. Several of these areas are domain specific, however,
the majority of them are not. We feel that the inclusion of
non-astronomical partnerships can provide tremendous insights.

\end{abstract}

\section{Introduction}

There is a data avalanche coming in astronomy, several large projects
are under way which will each produce many Terabytes of catalogued
data, covering a large part of the sky. Many of these catalogs will
cover different wavebands, from the X-rays to optical/infrared all the
way to the radio. We believe, that astronomy is about to undergo a
major paradigm shift, with data sets becoming larger and larger, and
more homogeneous, for the first time designed in the top-down
fashion. In a few years it will be much easier to `dial-up' a part
of the sky, when an astronomer needs a rapid observation than wait for
several month or a year to access a (sometimes quite small)
telescope. With the advent of inexpensive storage technologies and the
availability of high speed networks (vBNS, Abilene, Internet 2), the
concept of multiple, Terabyte size databases interoperating in a
seamless fashion is not an outlandish idea any more. More and more
catalogs will be added and linked to the existing ones, query engines
will be more sophisticated, and astronomers will have to be just as
familiar with mining data as with observing on telescopes.  Soon it
may become more cost effective to spend resources on integrating all
the data archives into a National Virtual Observatory, than building
another of the traditional observatories.

It is important to understand, that today's approaches of accessing
astronomical data do not scale well into the Terabyte regime - brute
force does not work! Let us assume a hypothetical 500 Gigabyte data
set. The most popular data access technique today is the World Wide
Web. Most university sites can receive data at the typical bandwidth
of about 15 kbytes/sec. The transfer time for this data set would be 1
year!  If the data is residing locally within the building, then
utilizing the common Ethernet access, typically at 1 Mbytes/sec, the
transfer time drops to 1 week.  If the astronomer is logged on to the
machine, which contains the data, all of it on hard disk, then using
the normal SCSI bandwidth it still takes 1 day to scan through the
data. We can conclude, that even faster hardware cannot support
hundreds of `brute force' queries per day, if no intelligent
solution is used. With Terabyte catalogs, customized data sets of a
few percent are still in the 10 GB range, thus an intelligent, high
level data management is needed.

Current efforts under way, which together will create a unified multi-
wavelength view of the Universe, include 2MASS, SDSS, POSS-2, FIRST,
COBE, MAP, GSC-II, ROSAT and GALEX, with wavelengths from X-rays
through UV, optical, near-IR to microwave and radio. Just the catalogs
listed here will provide all-sky (or almost all-sky information in
15 different bands. Their main characteristics are that they
have been designed to be homogeneous, are well calibrated and have a
good control over systematics. The catalogs will typically contain
over 100 million objects/pixels each, and will be multiple Terabytes
in size. The overlap between them will be substantial, but at the same
time there will be a lot of non-detections between different
archives. Once these catalogs are completed, their combination will
allow unique queries. 

Soon there will be an enormous pressure from the whole astronomical
community to integrate these separate archives into a seamlessly
interoperating entity. The public access of such amounts of data, the
necessary integration of the geographically distributed data sets and
their interoperability poses non-trivial problems, some of which are
technical, some are political and some are financial. Since most of
these archiving projects are just about to begin, it is not too late
to coordinate their architectures, but action must be taken very
shortly. New standards for the exchange of large amounts of data need
to be developed, and one has to address how to deal with the
distributed nature of the data. We also need to consider that the
computing platforms on which data are stored are heterogeneous, and
evolve rapidly. Political complications include the facts that
different data sets were created under the umbrella of different
federal agencies (NSF, NASA).

\section{Main Functions of the Virtual Observatory}

Today we understand, that this process of integrating ther archives is
inevitable, and within the next few years we will be faced with
concrete challenges. In this paper we address the next relevant issue,
what are the functionalities that these archives need to provide, and
what are the fundamental standards that we need to define before we
can continue on developing the tools.

First of all, the data in the archives must be maintained, curated.
We envisage that data would reside with the respective groups, who
know their own data best.  This of course means that the data is
scattered across the continent. The groups would maintain their own
data, and provide the storage, the documentation and the metadata,
describing the contents of the respective archives.

There are different types of users, whose support requires a widely
diverse array of resources.  Much of the general astronomy public will
only use these archives on a causal lookup basis, which will represent
a www interface, and a large number of rather simple queries. These
can be relatively easily supported via a central web-site with a state
of the art, but not too complex query engine. Intermediate users will
want to use the archive in a more elaborate fashion. Much more
difficult is to serve the power users, who would require multiple
searches in Terabytes of data, and extract hundreds of Gigabytes for
further processing. This task much resembles accessing supercomputer
resources, and our straw-man idea is to handle it like such.

We expect that most of the queries will be of exploratory nature, no
two queries will be exactly alike, at least for a while. Generally,
scientists will try to explore the multi-color properties of the
objects in the catalog, starting with small queries of limited
scope, then gradually making their queries more complex on a
hit-and-miss basis. Several typical types of activities need to be
supported: manual browsing, where one would look at objects in the
same general area of the sky, and manually/interactively explore their
individual properties, the creation of sweeping searches with complex
constraints, which extend to a major part of the sky, searches based
upon angular separations between objects on the sky,
cross-identifications with external catalogs, creating personal
subsets, and creating new "official" data products.

The data are inherently multidimensional, each object is represented by
several fluxes, position on the sky, size, redshift, etc.  Searching
for special categories of objects, like quasars, involves defining
complex domains in this N-dimensional space. Spatial relations will be
investigated, like finding nearest neighbours, or other objects
satisfying a given criterion within an angular distance. The output
size of the objects satisfying a given query can be so large that
intermediate files simply cannot be created. The only way to analyze
such data sets is to send them directly into analysis tools, thus
these will have to be linked to the archive itself.

The interconnection of the different sites could be accomplished via
one of the testbed programs connected to the next generation of
high-speed networks. The networking technology for the high speed
connectivity is here, but currently there few real-world applications,
which can generate a serious network load. The Virtual Observatory
could become a prime example of such creative network usage.

The creation and maintainance of the integrated framework involves
several components. First, one needs to understand the requirements.
Then appropriate standards need to be defined, then the tools
conforming to those standards, interfaces and protocols must be built.
Creation of the tools will require original development, by
implementing state of the art techniques from Computer Science. The
results will be very relevant much beyond astronomy, the whole society
is struggling with information retrieval. This development can only be
done by a wide collaboration, that involves not only astronomers, but
computer scientists and statisticians and even participants from the
IT industry.

\section{Discussion of the Common Standards}

\subsection{Communication Protocol}

Whenever people meet, they must agree (either implicitly or not) on a
protocol that they will use to share information (\ie a language).
This description also applies to the connection of disparate archives
and services, where the most basic requirement is that they must be
able to communicate amongst themselves.  This requirement can be
satisfied by defining a standardized communication protocol which the
individual components agree to employ.  Such a protocol goes beyond a
query language (which is a subset), and must provide support for
command functionality and performance monitoring, for both
inter-archive as well as archive-service communications.

The command functionality encompasses the needs for sharing metadata
and metaservices between archives and services, allow archives to
communicate amongst themselves, and monitor data streams (analysis
filtering).  In any complex, distributed archival system, the ability
to perform basic checkpoint operations on a query: stop, pause,
restart, and abort, is essential. Equally important is the ability to
provide feedback to the end-user (either a monitoring process or the
actual astronomer) on the status of the query.  An additional benefit
of a well-defined communication protocol is that pre-existing or
legacy archival services can be retrofitted (by mapping the new
standard onto existing services) in order to participate in
collaborative querying.

In a collaborative query model, a multi-wavelength search (\ie ``Find
all blue galaxies around dusty quasars'') will be distributed to
multiple archival centers. As a result certain optimizations can be
performed depending on the status of the archival connections (network
weather) in order to balance the resulting server load. Eventually, a
learning mechanism can be applied to analyze queries, and using the
accumulated knowledge gained from past observations (\ie artificial
intelligence), queries can be rearranged in order to provide further
performance enhancements.

A promising candidate for implementing the communication protocol is
the extensible markup language (XML). Different commands would be
encoded using different XML tags whose definitions are standardized
within a formal document type definition (DTD). The DTD would specify
the appropriate action for a given command, such as how to combine
multiple data streams and where the result should be sent.

\subsection{Hierarchical Metadata}

Continuing the previous analogy, the next step is to exchange relevant
descriptive information. For archives, this primarily consists of
their metadata (or data that describes the contents of the
archive). For services (which can be contained within an archive or
stand alone), it describes the specific nature of the work that the
service can perform (such as cross-identification or image
registration) and the expected format of the input and output
data. Depending on the need of the consumer, different amounts (or
levels) of detailed information might be required. For example, when
announcing its availability to other archival centers, an archive
might provide very general information concerning the specific
wavelengths of the data that it contains and the areal extants that it
covers. A specific user, however, might desire more detailed
information, such as the type of detectors used during the
observations, or the filter transmission efficiencies.

The nature and format of astronomical metadata clearly needs to be
standardized. A hierarchical format will enable users to ``drill
down'' until they reach the level of detail they require to perform
their analysis. We feel that this another area that XML might prove
beneficial. Not only is the hierarchical requirement satisfied, but it
is easily generated, and can be parsed by machines and read by humans
with equal ease. By adopting a standardized DTD, metadata can be
easily archived and accessed by any conforming application.

\subsection{High Performance Streaming Interface}

In the traditional exploration model where small numbers of
astronomical sources were analyzed, plain ASCII text was generally
sufficient for passing data between sites. With the advent of large
areal, digital surveys, however, the overhead of this approach
severely taxes the performance of a system. What is required is a
flexible system, in order to pass different types of data (\ie
tabular, spectral, or imaging data) that operates in a streaming
fashion (similar to MPI), so that analysis of the data does not need
to wait for the entire dataset before proceeding. As an example, we
might need to register images, or cross-identify sources from
different surveys.

A standardized binary data interchange format (the FITS--Flexible
Image Transport Standard) currently exists within the astronomical
community. The true efficacy of this format as a high performance
streaming interface is unclear. Another option is XML wrapped binary
transfers (which could employ a FITS approach).

\subsection{Efficient Object Interchange Format}

The primary purpose in federating multi-wavelength astronomical
surveys is to combine all available information for a set of
astronomical sources.  In order to analyze the data streaming from
multiple archives, a conceptual ``object'' identity is required for
each unique source (\ie the information that this source in archive A
is that source in archive B). As a result, a standard format for
efficiently interchange of astronomical ``objects'' is necessitated.

Traditionally this is accomplished by flattening the appropriate data
into a file (\eg the FITS Standard) and inflating the data at the
other end of the stream. However, this is not necessarily the best
method, and alternative techniques, such as serialization (using
either Java or C++) or an XML approach may prove more fruitful. The
adoption of a standard interchange format would shorten the
development cycle for user interface and analysis tools while
extending their applicability.

\subsection{Source Cross-Identification}

The cross-identification of billions of sources in both a static and
dynamic state over thousands of square degrees in a multi-wavelength
domain (Radio to X-Ray) is clearly an important and complex issue.
This ``join'' operation is accomplished by cross-identifying sources
in one archive with sources in multiple other archives. The process
is, of course, much more complicated than it initially appears, due
both to the nature of the data collection as well as the nature of
astronomical objects. Observational data is always limited by the
available technology, which varies greatly in sensitivity and
resolution as a function of wavelength (\eg optical instrumentation is
generally superioir to infrared instrumentation). The calibration of
the data (either spectral, temporal, or spatial) can also vary
greatly. As a result, it is often very difficult to unambiguously
match sources between different wavelength surveys; and , therefore,
an association often has an assigned probability and possibly more
attributes which solely describe the cross-identification object.

This problem becomes even more complicated when the large area of many
surveys, and the subsequent billions of objects, are
utilized. Furthermore, the facts that surveys have different limiting
flux sensitivities and that astronomical objects generally do not have
flat spectral shapes, implies that objects are often detected in one
wavelength region, might be much fainter, brighter, or even not
detected at all in another wavelength region (\ie sources can look
quite different depending on the wavelength region used). To overcome
this difficulty, a priori astronomical knowledge concerning the
spectral shape of ``known'' sources can be incorporated in order to
optimally determine source associations. A major concern with this
type of approach, however, is to prevent the misidentification of
novel sources. The limited spatial resolution of some surveys,
sometimes results in large positional error boxes, which can often
result in one-to-many or even many-to-many associations. The task is
also made more complicated by the areal extant of the
cross-identification (\ie full sky versus small areas). We also need
to be able to encapsulate the domain knowledge that was used to
determine the cross-identifications and their associated probability.

We see the need for standardization on two items: cross-identification
objects, and cross-identification generators. By defining a format for
encapsulating the astronomical knowledge that sources should be
associated, we can persist this newly generated relationship data and
begin to develop the often dreamed virtual observatories, where users
see a seamless multi-wavelength view of the universe. On the other
hand, some users will want to apply their own ideas to the problem of
cross-identification (perhaps incorporating new domain
knowledge). This will require a standardized interface for the new
association generators so that they might easily be incorporated into
existing frameworks (\ie a plug---and---play approach).

\subsection{Sky Partitioning and Navigation}

   The largest interoperability issue has to do with standards and
organization. As the volume of catalogued data in these archives will
approach a Terabyte, to ensure the integrity of the data it is prudent
to subdivide it into considerably smaller partitions, storing the
subdivisions as separate files. This partitioning should utilize
spatial information to a certain extent, something which has been done
by several existing surveys, either by simulating the POSS plates,
e.g. ROSAT, or utilizing their own partitons. Largely, these
subdivisions are hidden from the user, thus one could argue that it
does not matter how a catalogue is organized internally, as long as it
provides a common functionality. On the other hand this requires a lot
of parallel effort, and there are other bottlenecks on the horizon,
where a common subdivision scheme may offer a lot of benefits.
 The first point of common interest is the definition of common areas
over the sky, which can be universally used by astronomical databases.
The need for such a system is indicated by the widespread use of the
ancient constallations --- the first spatial index of the celestial
sphere. The existence of such an index, in a more computer friendly
form would substantially speed up cross referencing between future
catalogs, and would also enable a more consistent low-level
organization between different catalogs. The entire sphere would be
uniquely covered by these common regions. The boundaries of each area
and the area's nomenclature could be fixed for the foreseeable future.
A scheme for such subdivisions has been advocated earlier (Bartlett
1994).  Here we would like to adopt this idea, and carry it one step
further.

There is an obvious complication with this (or, for that matter, with
any) approach: the density of objects is not uniform over the sky.
Furthermore, the densities of different types of objects (and from
different types of surveys) follow different distributions. For
example, optical catalogs, primarily focusing on galaxies, like the
SDSS, will target areas away from the plane of the Galaxy. Infrared
surveys, like 2MASS, which cover the whole sky will have most of their
objects close to the Galactic plane. Generally, an efficient
partitioning of an archive will require an approximately balanced
distribution of the number of objects per partition, a strategy
already used by GSC-I. Having a common partitioning scheme, that
provides a balanced partitioning for all catalogs involved
simultaneously, may seem to be impossible, but as we will show below,
there is an elegant solution, a `shoe that fits all' --- by subdividing 
the sky in a hierarchical fashion!

\centerline{\epsfxsize=2truein\epsfbox{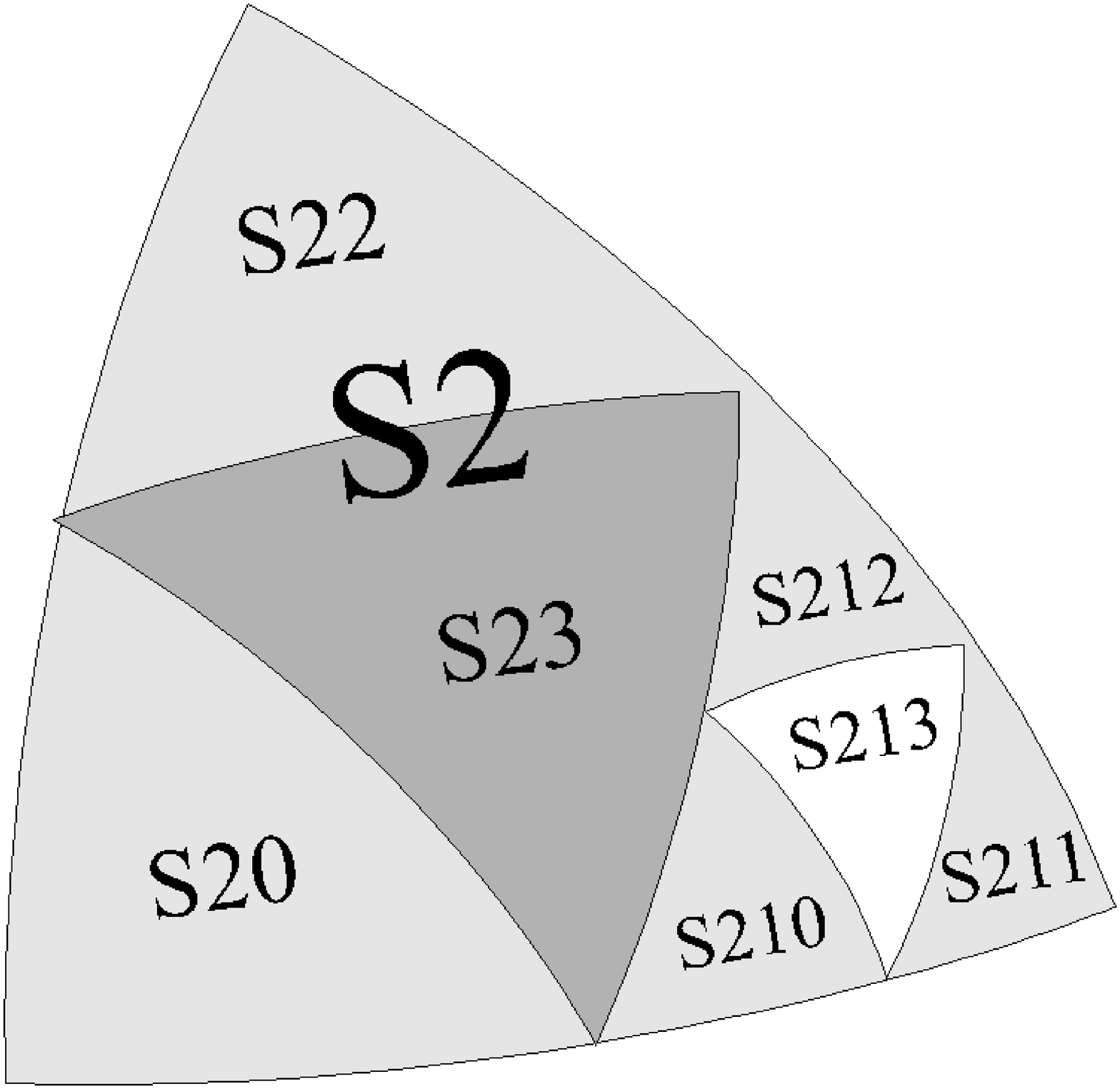}}

\noindent
{\bf Fig. 1.}
An illustration of how a spherical triangle S2 is subdivided
into four smaller ones. We also indicate how one of these, S21, is
subdivided further into S210...S213. Note, how the name keeps track
of the parents of a given region.
\bigskip

Our approach to dealing with this complication is the following:
instead of taking a given subdivision, we specify an increasingly
finer hierarchy of subdivisions, where each level is fully contained
within the previous one.  There would be a `base set' of selected
areas.  Each selected area can, if necessary, be divided into a number
(proposed to be 4) of approximately equal sub-areas.  Each sub-area can
be divided further into an additional 4 sub-areas, ad infinitum.
Such hierarchical subdivisions can be very efficiently represented on
the computer in the form of quad-trees (Samet 1990, Fekete 1990).

Each base selected area would be assigned a name, perhaps something
simple like S2 (S for South, 2 to identify one of the four). If an
archive has too many objects contained within S2 to handle
comfortably, it would divide S2 into sub-areas: S20, S21, S22, and
S23, for example.  If S21 contained too many objects, it could be
further subdivided into four more equal sub-areas designated
S210, S211, S212 and S213, etc.  This number, on how deep a
particular area would be subdivided in a given catalog, will be a
function of the storage size of the entry per object in the given
database.

This way, any of these selected areas are uniquely specified by their
name, even within the hierarchy -- the name will tell which area is
the parent, also how deep we are in the hierarchy. In such a way areas
in different catalogs map either directly onto one another, or one is
fully contained by another. 

For Cosmic Microwave Background experiments various similar pixelization
schemes have been proposed (Saff and Kuijlaars 1997, Greisen etal
1993, Tegmark 1996, Gorski 1997, Crittenden and Turok 1998). Mostly,
these schemes are not hierarchical, rather they create a grid of a
given resolution, which can then be optimized for various properties,
like having a lateral symmetry for fast spherical harmonic
computations, or equal area, usually at the expense of another
property, like similarity of shape.

One can create a hierarchical pixelization scheme starting from
an arbitrary Platonian solid, like tetrahedron, cube, dodecahedron
(Fekete 1990), or icosahedron (Tegmark 1996). Each of these have
emerged in the literature. The faces of these solids are regular
polygons, and their projections on the unit sphere is hierarchically
subdivided further, using great circles.  Such subdivision schemes are
well known in computer graphics (Ahn 1995).

Contrary to common wisdom, we propose to store the coordinates in a
Cartesian form, i.e. as a triplet of $x,y,z$ values per object.  The
$x,y,z$ numbers represent the position of objects on the unit sphere.
While at first this may seem to increase the required storage (three
numbers per object vs two angles), it makes querying the database for
objects within certain areas of the celestial sphere, or involving
different coordinate systems considerably more efficient. Again, this
is by no means new, this was used highly successfully by GSC-I.

The coordinates in the different celestial coordinate systems
(Equatorial, Galactic, Supergalactic, etc) can be constructed from the
Cartesian primitive, on the fly.  Conversions are performed by
standard functions which convert between the angular and cartesian
coordinate objects. This approach allows the conversion to be
specified at run time, allowing recalibrations to be seamlessly
integrated.

The most common types of queries involving celestial coordinates are
of the four fundamental types listed below. As we will see, having
our coordinates stored as three-dimensian vectors will make the evaluation
of all these queries considerably easier. Consider the fact, that
all objects above a given latitude in a certain coordinate system can
also be looked at as the objects on one side of the plane in 3d space,
which corresponds to the small circle of the latitude. The four fundamental
query types are the following:

\begin{itemize}
\smallskip\begingroup\parskip=0pt

\item all objects within a certain spherical distance of a given
point. These can be recast into intersections on the unit sphere with a plane,
with the plane normal pointing at the center. 

\item objects above a given latitude in any spherical coordinate
system are above a three dimensional plane, intersecting the sphere 
at that latitude circle.

\item any great circle, thus any longitude is represented by the
intersection of the sphere with a plane going through the origin.

\item Boolean combinations (AND, OR, NOT) of the above
\endgroup
\smallskip
\end{itemize}

All these primitive queries are so-called half-space queries, where
each of the constraints excludes half of the three dimensional space
(but not half the sky!). Evaluating these queries is very inexpensive
in terms of CPU cycles, since they all involve a computation of a
linear combination rather than a spherical distance for example. Also,
even though the query may contain constraints specified in several
different coordinate systems, these coordinates need not be stored or
even evaluated for each object, only linear combinations of the
Cartesian are evaluated for every constraint. In this way, every
spherical coordinate system is handled in the same fashion.  Besides,
these linear combinations with their Boolean operations can trivially
be represented in query languages like SQL or OQL.


\centerline{\epsfxsize=4.1truein\epsfbox{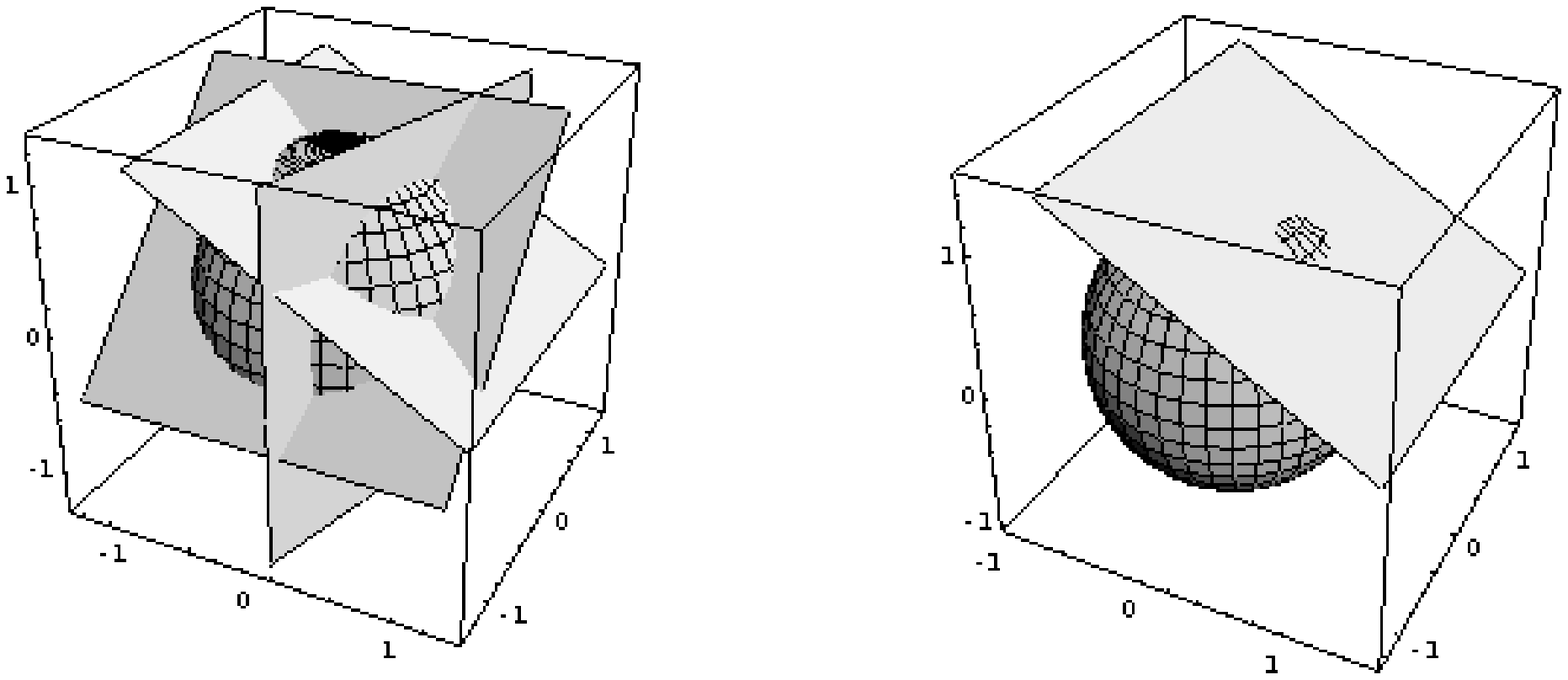}}

\noindent{\bf Fig. 2.}
Schematic representation of how the 3-dimensional Cartesian 
representation of coordinates make queries like objects within a
certain spherical distance to a given point, or combination of 
constraints in multiple coordinate systems become particularly 
simple, since they correspond to testing linear combinations
of the Cartesian coordinates instead of complicated trigonometric
expressions.
\bigskip

\subsection{Uniform Handling of Detections and Upper Limits}

The advantage of using a logarithmic scale to measure astronomical
fluxes is obvious, the magnitude scale is able to span a huge dynamic
range. When relative colors are needed, they can be computed by
differencing magnitudes measured in different bandpasses. These
advantages are quite clear for bright objects, where noise is not an
issue. On the other hand, as fluxes become comparable to sky- and
instrumental noise, the magnitudes become more and more scattered due
to the singularity in the magnitude scale at zero flux. The cases,
where due to a downward sky fluctuation we measure negative flux for
an object cannot be represented with magnitudes. People have generally
handled these cases by specifying detection flags, and encoding the
negative flux somehow. These problems become even more pronounced as
we work in multicolor space. An object can be well detected and
measured in several of the bands, and still fail to provide a
measurable flux in others. This information of non-detection is
meaningful, beyond just a detection flag, in flux space the object
would be represented as a multi-variate Gaussian probability centered
on the flux values, not necessary all positive. In magnitude space,
the error ellipsoid of such an object would have an infinite extent in
some of the directions, making meaningful multicolor search criteria
in a database extremely complex, if not impossible. All the
non-detections in one or more bands have to be isolated, and treated
separately. One can conclude, that most problems arise from the
singular behaviour of the logarithm function. We will discuss here a
suggestion by Jim Gunn and Robert Lupton, to keep all the advantages
of magnitudes, without their disadvantage. We expect that this will be
the way how fluxes will be represented in the SDSS Science Archive.

We propose to use the inverse hyperbolic sine function instead of the
logarithm, in defining a new magnitude scale. This function behaves as
the logarithm for large values of its arguments, while staying linear close
to zero.
\begin{equation}
    \sinh^{-1} (x) = \ln \Bigl[ x + \sqrt{x^2+1} \Bigr] \rightarrow 
	\cases{\ln 2x, & if $x\gg 1$\cr x, & if $|x|<1$\cr}
\end{equation}
Let us consider the dimensionless, normalized flux $x$, with a correct
zero-point, defined as $x=f/f_0$. The usual apparent magnitude $m$ can
be written as

\begin{equation}
	m = -2.5 \log_{10} x = -2.5/\ln 10 \ln x = -a \ln x.
\end{equation}

\noindent We can also define the new magnitude $\mu$ as
\begin{equation}
	\mu(x) = -a \Bigl[ \sinh^{-1} \Bigl( {x\over 2b}\Bigr) 
		+ \ln b\Bigr]
\end{equation}
Here $a$ and $b$ are constants, $a=2.5/\ln 10=1.08574$, and $b$ is an
arbitrary 'softening', set to the optimum value of $b= \sqrt{a}\sigma
= 1.042\sigma$, where $\sigma$ is the typical error in a object's flux.
The value of $b$ determines at which flux level will the linear
behaviour set in.  Consider the asymptotic behaviour of $\mu$, for
both high and low $x$:
\begin{equation}
	\lim_{x\to\infty}{\mu(x)} = -a \ln x = m 
		\qquad\qquad\qquad
	\lim_{x\to0}{\mu(x)} = -a \Bigl[{ x\over 2b} + \ln b\Bigr].
\end{equation}
Thus for $x\rightarrow\infty$, $\mu$ approaches correctly $m$, for any
choice of $b$. Now consider the low $x$ asymptotic behaviour, and see
that it is indeed linear in $x$, but this behaviour does depend on $b$.

We have to take into account, that all the flux measurements will have
an error. Differences which are much smaller than the flux error do
not matter much. Assume that the normalized flux $x$ has a Gaussian
error $\sigma^2=\avrg{x^2}-\avrg{x}^2$.  Let us compute the variances
for the two magnitudes and their difference. The arrows indicate the
asymptotic behaviour for $x\rightarrow0$.
\begin{eqnarray}
	\avrg{\Delta m^2}\  =&\  {a^2\sigma^2\over x^2}
	\rightarrow\quad &{a^2\sigma^2\over x^2}		\cr
	\avrg{\Delta\mu^2}\ =&\  {a^2\sigma^2\over 4b^2 + x^2} 	
	\rightarrow\quad &{a^2\sigma^2\over 4b^2} 		\cr
	m-\mu\ =&\  a \ln \Bigl[{1 + \sqrt{1+4b^2/x^2}\over 2}\Bigr]
	\rightarrow\quad & -a \ln({x\over b})
\end{eqnarray}

In the discussion above, we considered the idealized case, that all
objects have the same Gaussian error, mostly dominated by sky noise.
This of course assumes, that all the apertures in the survey are the
same. Also, the actual noise may vary with the position on the sky,
like the vicinity of a bright object, etc. Therefore, we need to
distinguish between the {\it nominal\/} and {\it actual\/} scatter in
a given object's flux. The {\it nominal\/} dispersion is the one
averaged over all objects in the survey, dominated by the faintest
objects at the edge of detection. This quantity is an average one.
The {\it actual\/} dispersion is arising indeed from local noise,
varying aperture sizes, etc. The two will only differ for relatively
big and bright objects, or for small objects sitting on the wings of 
bright stars. 

\centerline{\epsfxsize=4.1truein\epsfbox{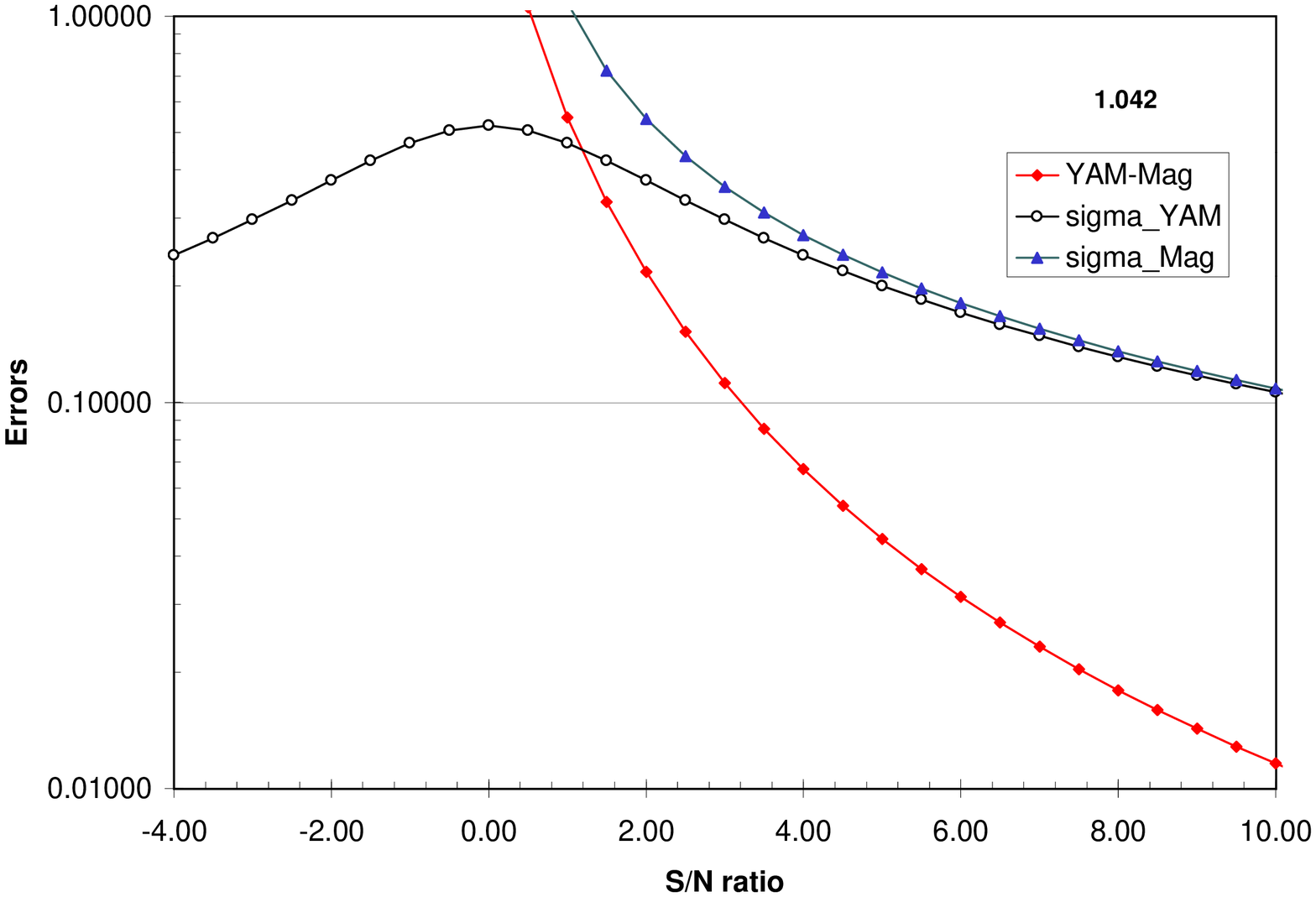}}

\noindent{\bf Fig. 3.}
We show the variances and the difference as a function of the
signal-to-noise $x/\sigma$. The open circles represent $\Delta\mu$,
the filled triangles represent $\Delta m$ and the filled diamonds
correspond to $m-\mu$. The figure corresponds to the optimal choice of
$b=1.042$, as discussed below. Note, how small the deviation $m-\mu$
is compared to the magnitude error. For a S/N of 3, the difference is
less than 0.1, while $\Delta m=0.3$.
\bigskip


We suggest, that one global choice of $b$ be adopted for each of the 
filters for a given survey, which
correspond to the mean {\it nominal\/} dispersion of the faintest
objects in each band.  Then the actual fluxes can be recovered with a
global inverse transformation as needed. Photometric zero points can
still be handled as usual, since they occur at such a high flux level,
that the difference between the two magnitudes is a fraction of a
percent. Multicolor searches would then deal with objects, which have
well-localized error ellipsoids. The error ellipsoids can follow the
{\it actual\/} dispersion, which can then exceed the nominal maximum
value of 0.52. If an object is not detected in a given band, its error
ellipsoid (of about 0.52 magnitudes) would be centered on the value 
corresponding to zero flux, but the object would be still very well
localized.

\subsection{Advanced Query Language}

In typical astronomical queries the users fill out a www form, with 
minimal constraints, like a point on the sky and a search radius, or a
name of a well known object, or a right ascension and declination range,
combined with a magnitude range. For more complex science, in particular
for finding interesting objects with unusual colors, one needs much more
complex constraints. In the process of classifying different types of 
objects, typically rather complex decision surfaces are needed, most often
in a higher dimensional space. Like in order to find quasars, we need objects
which are point-like, and have unusual colors, which can be both very blue
or very red, depending on the redshift of the quasars. Such decision surfaces
are very complicated. How can these be represented? We would like to promote
the idea, that by allowing linear combinations of magnitudes, one can create
a multidimensional half-space query. Their Boolean combinations, similar to our
ideas for the sky navigation, are multidimensional polyhedra. These can be
expressed in the more advanced formal query languages, like SQL or OQL.

There are of course queries, which will be required for advanced
astronomical searches, which involve spatial relations, proximities
between objects.  These cannot be expressed by traditional SQL
expressions, we need new extensions to the language. These extensions
should contain a spherical distance operator between two objects, and
a distance based on a Eucledian metric in magnitude space. Such an
operator needs to have not only a proximity radius in magnitude space,
but a description of the subspace the distance should be calculated
in. Such a query could be: {\it Find me all galaxies, which have
another one within 10 seconds of arc, with colors within 0.2
magnitudes in $r$ and $b$}. The ability of embedded SELECT statements
is a necessary condition to be able to describe the complex
crieteria. Thus, we envisage that the astronomy community will create
a small number of common extensions to SQL, describing now functions
with a spatial content, and most of the archives will adopt these.

Searching distributed archives with a single query statement is much
more complex. One can possibly accommodate this by 
\begin{itemize}
	\item either using such simple queries, that only relate to attributes
common to all archives
	\item include specific scope statements in the FROM clauses relevant
for each of the archives, by keeping track of the individual data models
	\item by creating a federated data model, that is the union of all the data models in the Virtual Observatory, and preprocess the query using this knowledge
	\item via an intelligent federated layer, that given a query, interrogates the archives about their contents and distributes the query accordingly
\end{itemize}
This is an area that will require a lot of further research.

There will also be queries which will only return aggregate
quantitites, like tyhe number of objects with a given property, or the
average of an attribute, given a set of constraints. Such queries may
be executed very efficiently if we have a condensed representation of
the data. Condensed representations enable one to retrieve such
aggregate information without performing the detailed query, and are an
active area of modern computer science (Moore 1998).

\subsection{Multiepoch observations}

Many astronomical objects are variable, and observations made at different
times may thus give different results about their brightness. Other objects,
like asteroids are mnoving rather rapidly on the sky. It is important to keep
track of the information when the respective observations were made. It should
be easy to extract all observations of a given object, in all archives, 
ordered in sequence. One can then develop various hypotheses about the nature
of their variability (like what is the significance that this is an RR Lyrae
variable). There may be several hypotheses equally likely. These qould be
stored in a database. As further observations become available, one can
reexamine and update these hypotheses, refine their parameters.

A similar approach could be done with asteroids --- from their first few observations one can already calculate a crude trajectory. As at another time a new
asteroid is observed, first one should check, whether it is consistent with 
any of the previously detected asteroids, by brightness and trajectory, and
if so, one can refine those models. If no match is found, a new hypothesis
is calculated and inserted into the database.

On the time series one can create increasingly refined models for
certain classes of objects, like variable stars. One can eventually
extract light curves, possibly in multicolor, or tag single transient
events like supernovae. This is an area of astronomy which is in rapid
development, but most of these queries have been restricted to
datasets covering a small part of the sky.

\section{Summary}

In this paper we listed, without the sake of completeness some of the
important problems facing the next generation archives in astronomy.
The future lies in their seamless integration, but there is lot to be
done, the Virtual Observatory may be just as complex if not more so
than a traditional one. However, astronomy will be very different ten
years from now, much more research will be done via the integrated archives,
and data mining will be an integral part of every astronomer's daily routine,
just as using a web browser and email is today.

\acknowledgments

We wish to thank George Djorgovski, John Good, Gretchen Greene, Barry
Lasker, Tom Prince, Doug Reynolds, Gyula Szokoly, Ani Thakar, and Roy
Williams, for stimulating discussions.

\end{document}